\documentclass[print]{article}
\usepackage{amsmath,amssymb}
\usepackage{graphicx}
\newcommand{\lvec}[1]{|#1\!\!>}

\begin{document}

\section*{Soliton Atom Laser with Quantum State Transfer Property}

\begin{center}Xiong-Jun Liu$^{a,b}$\footnote{Electronic address:
phylx@nus.edu.sg}, Hui Jing$^{c}$\footnote{Electronic
address:jinghui@wipm.ac.cn} and Mo-Lin Ge$^{b}$\end{center}

\begin{center}a. Department of Physics, National University of
Singapore, 10 Kent Ridge Crescent, Singapore 119260, Singapore \\
b. Theoretical Physics Division,
Nankai Institute of Mathematics,Nankai University, Tianjin 300071, P.R.China\\
 c. State Key Laboratory of Magnetic Resonance and Atomic and Molecular
 Physics,\\
 Wuhan Institute of Physics and Mathematics, CAS, Wuhan 430071, P. R.
 China\end{center}

\begin{abstract}

We propose a scheme to obtain soliton atom laser with nonclassical
atoms based on quantum state transfer process from light to matter
waves in nonlinear case, which may find novel applications in,
e.g., an atom interferometer. The dynamics of the atomic gray
solitons and the accompanied frequency chirp effect are discussed.
\\

PACS numbers: 03.75.-b, 42.50.Gy, 03.65.Ta, 03.67.-a
\end{abstract}

\baselineskip=13pt

\indent Since the first pulsed atom laser was created in 1997
through RF output coupling of a trapped atomic Bose-Einstein
condensate \cite{atom laser}, there have been many interests in
preparing a continuous atom laser and exploring its potential
applications in, e.g., gravity measurements through atom
interferometry \cite{interferometer}. Although a sub-quantum-noise
atom laser is expected to be crucial to improve the interferometer
sensitivities, the difficulties for the atomic beam to propagate
over a long distance heavily restrict its actual performance
\cite{noise}. Some time ago Drummond et al. proposed to use
mode-locking technique to stabilize the atom laser based on the
generation of a dark soliton in a ring-shaped condensate
\cite{Drummond}. Other related works are the atomic soliton
formation and its stationary transmission in a travelling optical
laser beam \cite{Zhang} or a waveguide \cite{Leboeuf} for a dense
atomic flow. While optical techniques are by now well developed to
make and control a soliton laser \cite {Solitons} (even using an
interferometer \cite {Opinterferometer}), less progress has been
made for ultracold atoms. We here propose a novel scheme to obtain
a soliton atom laser with nonclassical characteristic via the
versatile quantum state transfer technique \cite{1,2,3,4}.

Recently, by manipulating two external lights for an ensemble of
$3$-level $\Lambda$ type atoms, the physical mechanism of
Electromagnetically Induced Transparency (EIT) \cite{5} has
attracted much attention in both experimental and theoretical
aspects \cite{1,2,3}, especially after the dark-states polaritons
(DSPs) theory \cite{4} was proposed and thereby the rapid
developments of quantum memory technique, i.e., transferring the
quantum states of photon wave-packet to collective Raman
excitations in a loss-free and reversible manner. By extending the
transfer technique to matter waves, a wonderful scheme was
proposed to make a continuous atomic beam with nonclassical or
entangled states \cite{6}, which was later confirmed also for
double-$\Lambda$ 4-level atoms \cite{7}.

Nonetheless, in these pioneering works the role of nonlinear
atomic interactions in the quantum state transfer process was
neglected \cite{6,7}. In this paper, by studying the quantum
states transfer technique from photons to atomic beam in the
nonlinear case, we propose a scheme to obtain a soliton atom laser
with nonclassical atoms. The dynamics of present gray-solitons is
shown to be free of the atomic frequency chirp effect occurring in
state transfer process. Essentially being different from output
couplings of atomic solitons formed in a trapped condensate
\cite{Drummond,9,10,11}, this scheme should be realizable in the
next generation of experiments.

The development herein is outlined as follows. Firstly, we study
the role of nonlinear atomic interactions in the transfer process
of quantum states from probe field to matter waves in adiabatic
condition. For present purpose, the collisions between the
initial-state atoms ($\Phi_1$) are omitted by, e.g., applying the
recently developed novel technique of magnetic-field-induced
Feshbach resonance \cite{Feshbach}, while nonlinear interaction
between the generated atoms ($\Phi_2$) is well considered. The
quantum transfer character is still confirmed except for an
additional phase leading to the frequency chirp effect
\cite{chirp}. Secondly, we focus on the formation and dynamics of
atomic solitons in the output beam, including the speed, the
free-chirp property, etc. Finally, we discuss the possibilities of
further manipulations of soliton atom laser.

Turning to the situation of Fig.1. The situation we consider is a
beam of three-level $\Lambda$ type atoms moving in the $z$ direction
interact with a quantum probe and a control Stokes field and the
former field is taken to be much weaker than the later. Atoms in
different internal states are described by three bosonic fields
$\hat\Psi_{\mu}(z,t) (\mu=1,2,3)$. The Stokes field coupling the
transition from meta-stable state $|2\rangle$ to excited one
$|3\rangle$ can be described by the Rabi-frequencies
$\Omega_s=\Omega_{0}(z)e^{-i\omega_{s}(t-z/c')}$ with $\Omega_0$
being taken as real, and $c'$ denoting the phase velocities
projected onto the $z$ axis. The quantized probe field coupling the
transition from ground state $|1\rangle$ to $|3\rangle$ are
characterized by the dimensionless positive frequency component
$\hat E^{(+)}_p(z,t)=\hat{\cal E}(z,t)e^{-i\omega_{p}(t-z/c)}$. We
can introduce the slowly-varying amplitudes
$\hat\Psi_1=\hat\Phi_1e^{i(k_0z-\omega_0t)}$,
$\hat\Psi_2=\hat\Phi_2e^{i[(k_0+k_p-k_s)z-(\omega_0+\omega_p
-\omega_s)t]}$ and
$\hat\Psi_3=\hat\Phi_3e^{i[(k_0+k_p)z-(\omega_0+\omega_p)t]}$, where
$\hbar\omega_0=\hbar^2k^2_0/2m$ is the corresponding kinetic energy
in the average velocity, $k_p$ and $k_s$ are respectively the vector
projections of the probe and Stokes fields to the $z$ axis. The
atoms have a narrow velocity distribution around $v_0=\hbar{k_0}/m$
with $k_0\gg|k_p-k_s|$, and all fields are assumed to be in
resonance for the central velocity class. The Hamiltonian of the
total system is $H=H_0+H_{coll}+H_I$, where (under the $s$-wave
approximation)
\begin{eqnarray}\label{eqn:Hamiltonian}
H_0&=&\sum_{j=1,2,3}\int
dz\hat\Psi_j^{\dag}(-\frac{\hbar^2}{2m}\frac{\partial^2}{\partial
z^2}+\hbar V_j)\hat\Psi_j,\\
H_1&=&\sum_{i,j=1,2}\hbar U_{ij}\int dz\hat\Psi^{\dag}_i
\hat\Psi^{\dag}_j\hat\Psi_i\hat\Psi_j,\\
H_I&=&-\int dz\hat\Psi^{\dag}_3[\hbar g \hat E(z,t)]\hat\Psi_1-\int
dz\hat\Psi^{\dag}_3[\hbar\Omega_s(z,t)]\hat\Psi_2+h.c.
\end{eqnarray}
are the free atomic part, atomic collision part and atom-field
interaction part, respectively. The Heisenberg equations for the
bosonic field operators are governed by \cite{eq}
\begin{eqnarray}\label{eqn:1}
i\hbar\frac{\partial\hat\Psi_1}{\partial
t}=[-\frac{\hbar^2}{2m}\frac{\partial^2}{\partial z^2}
+V_1(z)+U_{11}\hat\Psi_1^{\dag}\hat\Psi_1+U_{12}\hat\Psi_2^{\dag}\hat\Psi_2]\hat\Psi_1
+\hbar g\hat E_p^{\dag}\hat\Psi_3
\end{eqnarray}
\begin{eqnarray}\label{eqn:2}
i\hbar\frac{\partial\hat\Psi_2}{\partial
t}=[(\epsilon_{12}-\frac{\hbar^2}{2m}\frac{\partial^2}{\partial
z^2})
+V_2(z)+U_{21}\hat\Psi_1^{\dag}\hat\Psi_1+U_{22}\hat\Psi_2^{\dag}\hat\Psi_2]\hat\Psi_2
+\hbar\Omega_s^*\hat\Psi_3
\end{eqnarray}
\begin{eqnarray}\label{eqn:3}
i\hbar\frac{\partial\hat\Psi_3}{\partial
t}=[(\epsilon_{13}-\frac{\hbar^2}{2m}\frac{\partial^2}{\partial
z^2})+i\hbar\gamma+V_3(z)]\hat\Psi_3 +\hbar g\hat
E_p\hat\Psi_1+\hbar\Omega_s\hat\Psi_2
\end{eqnarray}
where $V_i(z) (i=1,2,3)$ are the longitudinal external effective
potentials of which $V_1(z)$, similar to the previous works, will be
chosen as $V_1(z)=0$ in the following derivation
\cite{6,7,potential}, $g$ is the atom-field coupling constant
between the states $\lvec{1}$ and $\lvec{3}$ \cite{4},
$\epsilon_{13}=\hbar(\omega_{31}-\omega_{p})$ and
$\epsilon_{12}=\hbar(\omega_{21}-\omega_{p}-\omega_{s})$ are
energies of the single and two-photon detunings. $\gamma$ denotes
the loss rate out of the excited state and the scattering length
$a_{ij}$ characterizes the atom-atom interactions via
$U_{ij}=4\pi\hbar^2a_{ij}/m$. Since almost no atoms occupy the
excited state $\lvec{3}$ in the dark-state condition fulfilled in
the EIT technique, the collisions between $\lvec{3}$ and lower
states can be safely omitted. The propagation equation of the probe
field reads: $
(\frac{\partial}{\partial{t}}+c\frac{\partial}{\partial{z}})\hat{\cal
E}(z,t) =-ig\hat\Psi^{\dag}_1\hat\Psi_3e^{-i(k_pz-\omega_pt)}. $ And
the depletion of the strong classical Stokes field is neglected.

We can study the adiabatic situation by ignoring the two
photon-detuning and the decaying of excited states. From
Eq.(\ref{eqn:3}), we obtain that $\hat\Psi_2=-\frac{g\hat
E(z,t)}{\Omega_s(z,t)}\hat\Psi_1=-\frac{g\hat{\cal E}(z,t)}
{\Omega_0(z)}\hat\Psi_1e^{i[(k_s-k_p)z-(\omega_s-\omega_p)t]}$.
Consider a stationary input of atoms in state $\lvec{1}$ and in the
limit of weak probe field, we have \cite{6,7}
$\langle\hat\Psi_2^{\dag}\hat\Psi_2\rangle\ll\langle\hat\Psi_1^{\dag}\hat\Psi_1\rangle$
and then ignore the depletion of the ground-state atoms and the
nonlinear term involving $\hat\Psi_2$ in Eq.(\ref{eqn:1}).
Furthermore, we may choose a zero-value scattering length $a_{11}$
through the Feshbach resonance technique \cite{Feshbach}. Hence the
depletion of atoms in state $|1\rangle$ and all the nonlinear terms
in Eq.(\ref{eqn:1}) can be ignored and the solution can be written
as $\hat\Psi_1(z,t)\approx\langle\hat\Psi_1\rangle=n
e^{i(k_0z-\omega_0t)}$ \cite{6}, where $n$ is the constant total
density of atoms. The $\hat\Psi_3$-field reads:
$\hat\Psi_3(z,t)=-\frac{1}{\hbar\Omega_s}[\frac{\hbar^2}{2m}\frac{\partial^2}{\partial
z^2} +i\hbar\frac{\partial}{\partial
t}-V_2(z)-U_{21}n-U_{22}\hat\Psi_2^{\dag}\hat\Psi_2^2]\frac{g\hat
E(z,t)}{\Omega_s}\Psi_1 $. Now we reach the following equation of
motion for the radiation field:
\begin{eqnarray}\label{eqn:elm}
i\hbar\bigl[(1+\frac{g^2n}{\Omega^2_0(z)})\frac{\partial}{\partial{t}}
+c(1+\frac{g^2n}{\Omega^2_0(z)}\frac{v}{c}-i\frac{g^2n}{\Omega^2_0(z)}
\frac{v}{c}\frac{\partial_z\ln\Omega_0}{k})
\frac{\partial}{\partial{z}}\bigl]\hat{\cal E}(z,t)\nonumber\\
-\bigr(A(z)+B(z)\hat{\cal E}^{\dag}\hat{\cal E}\bigr)\hat{\cal
E}(z,t)
=i\hbar\frac{g^2n}{\Omega^2_0(z)}v(\frac{\partial}{\partial{z}}
\ln\Omega_0(z))\hat{\cal E}(z,t).
\end{eqnarray}
Here we neglect the second derivative of slowly-varying amplitude
$\hat{\cal E}$ and assume the sufficiently slowly spatial
variations of $\Omega_0$ \cite{6,7}.
$A(z)=\frac{g^2n}{\Omega_0^2}\bigl[(V_2(z)+nU_{21})+\frac{v_0^2}{2k_0^2}\bigl((\frac{\partial\ln\Omega_0}{\partial
z})^2-\frac{\partial^2\ln\Omega_0}{\partial z^2}\bigl)\bigl]$,
$B(z)=U_{22}\frac{g^4n^2}{\Omega^4_0(z)}$, $k=k_0+k_p-k_c$ and
$v=v_0+v_r$ with $v_r=\hbar(k_p-k_c)/m$ being the recoil velocity
for $|1>\rightarrow|3>$ transition in $z$ direction. Although the
probe field is weak (and then the generated atomic beam $\Phi_2$
is also weak, see the following context), the ultra-slow light
case ($\frac{g^2n}{\Omega^2_0}\gg1$) can greatly enhance the
nonlinear interaction. Since $k_0$ is a very large factor $
(k_0\gg|k_p-k_c|)$, we have $|\partial_z\ln\Omega_0/k|\ll 1$. For
this the corresponding parts in above equation can safely be
neglected. By introducing the mixing angle $\theta(z)$ according
to $\tan^2\theta(z)\equiv\frac{g^2n}{\Omega_0^2}\frac{v}{c}$, we
can obtain the final solution
\begin{eqnarray}\label{eqn:6}
\hat{\cal E}(z,t)=\frac{\cos\theta(z)}{\cos\theta(0)}\hat{\cal
E}\bigl(0,T\bigl)\exp\bigl(-i\hat{\cal E}^{\dag}(0,T)\hat{\cal
E}(0,T)\int_0^z|\frac{\cos\theta(\xi)}{\cos\theta(0)}|^2B'(\xi)d\xi-i\int_0^zA'(\xi)d\xi\bigl)
,
\end{eqnarray}
where $T=t-\tau(z)=t-\int^z_0{d\xi V^{-1}_g(\xi)}$ is the time scale
in rest frame, the group velocity
$V_g=c(1+\frac{g^2n}{\Omega^2}\frac{v}{c})/(1
+\frac{g^2n}{\Omega^2})$ approaches $v$ for $\Omega(z)\rightarrow0$.
$A'(\xi)=A(\xi)/(1+\frac{g^2n}{\Omega_0^2(\xi)})\hbar$ and
$B'(\xi)=B(\xi)/(1+\frac{g^2n}{\Omega_0^2(\xi)})\hbar$. In
particular, by assuming $\theta(0)=0$ and $\theta(L)=\pi/2$ at the
input and output regions respectively, one clearly sees that the
slowly-varying amplitude of the bosonic field $\hat\Psi_2$ can be
written as
\begin{eqnarray}\label{eqn:7}
\hat\Phi_2(z,t)=\sqrt{\frac{c}{v}} \ \hat{\cal
E}\bigr(0,t-\tau(z)\bigr)\exp\bigr(i\Delta\phi(z,t)\bigr), \ \
(z\geq L)
\end{eqnarray}
with an additional quantum phase
$\Delta\phi(z,t)=\int_0^zA'(\xi)d\xi\bigl)+\hat{\cal
E}^{\dag}(0,T)\hat{\cal
E}(0,T)\int_0^z|\cos\theta(\xi)|^2B'(\xi)d\xi$ [15]. The factor
$\sqrt{c/v}$ shows that the input light propagate with velocity
$c$ while the output atoms with $v$. Obviously, the presence of
intrinsic nonlinear atomic interaction leads to a self-phase
modification (SPM) in the quantum state transfer process. Due to
the time-dependent character of $\Delta\phi$, it indicates an atom
laser with frequency chirp
$\delta\omega=-\frac{\partial\Delta\phi(z,t)}{\partial
T}=-\frac{\partial|{\cal E}(0,T)|^2}{\partial
T}\int_0^z|\cos\theta(\xi)|^2B'(\xi)d\xi$, and $|{\cal
E}(0,T)|^2=\langle\hat{\cal E}^{\dag}(0,T)\hat{\cal
E}(0,T)\rangle$, which may hold the promise to find some actual
applications \cite{chirp}.

As a concrete example, let us consider a Gaussian or super Gaussian
envelop for the input probe field (see Fig.2(a)) and then obtain the
following expression:
\begin{eqnarray}\label{eqn:chirp}
\delta\omega(T)=\frac{2m}{T_0}\int_0^z|\cos\theta(\xi)|^2B'(\xi)d\xi
\biggr(\frac{T}{T_0}\biggr)^{2m-1}
\exp\biggr[-\biggr(\frac{T}{T_0}\biggr)^{2m}\biggr], \ \ (z\geq L)
\end{eqnarray}
where $T_0$ is the normalized time scale, $m=1$ and $m=3$
characterizes the Gaussian and super Gaussian pulse, respectively.
Using
$\Lambda_0(z)=\frac{2}{T_0}\int_0^z|\cos\theta(\xi)|^2B'(\xi)d\xi$
to normalize the frequency, the chirp development is shown in
Fig.2(b), from which one can clearly see that the frequency chirp of
the output atom laser is significantly dependent on the gradient of
the input probe pulse's front and tail.

Now we proceed to reveal the formation of atomic solitons in the
output beam and probe some of its novel properties. For
simplicity, our following investigations will adopt the mean-field
description (or quasi one-dimension Gross-Pitaevskii equation)
\cite{GPG}. Note that the above quantum state transfer process is
clearly independent on this approximation. For this the bosonic
field $\hat\Psi_2(z,t)=\Psi_2(z,t)+\hat\psi_2$, where the
condensate wave function $\Psi_2$ satisfies the nonlinear
Schr\"{o}dinger equation (NLSE) which supports a gray- (or
bright-) soliton solution for a positive (or negative) scattering
lenght $a_{22}$ and $\hat\psi_2$ denotes the quantum fluctuation.
Taking into account of the mixing angle $\theta=\pi/2$ in the
region $z\geq L$, or the Rabi frequency of the Stokes control
field $\Omega_s\rightarrow0$ (meanwhile no transition between the
atoms in state $|1\rangle$ and $|2\rangle$), we obtain motion
equation for $\Psi_2$ in the following form
\begin{eqnarray}\label{eqn:8}
i\hbar\frac{\partial\Psi_2}{\partial
t}=-\frac{\hbar^2}{2m}\frac{\partial^2\Psi_2 }{\partial z^2}
+V_{2eff}(z)\Psi_2 +U_{22}|\Psi_2|^2\Psi_2
\end{eqnarray}
with an effective potential $V_{2eff}=V_2+nU_{21}$. For positive
scattering length $a_{22}$ or repulsive atomic interaction, the
solution of above equation describing a gray soliton moving in
propagating background wave function is \cite{9,10,11} $:
\Psi_2(z,t)=\Psi^0_2(z,t)\bigr\{i\sqrt{1-\eta^2}+\eta\tanh\bigl[\frac{\eta}{\sqrt{\beta}}(z-z_0(t))\bigl]\bigr\}
$, where $\beta=1/(8\pi|\Psi_2|^2a_{22})^{1/2}$, the external trap
$V_2(z)$ is chosen such that $V_{2eff}=0$ \cite{6,7,potential},
the background wave function
$\Psi^0_2(z,t)=\Phi_2(z,t)\exp\bigl[i(k z-\int^t_{t_0}\lambda
dt'\bigl)+i\varphi_0\bigl]$ with
$\lambda=\frac{k^2}{2m}+\frac{U_{22}}{\hbar}|\Phi_2|^2$ and
$\Phi_2=\langle\hat\Phi_2\rangle$, the centra position of the
soliton at time $t$ is $z_0(t)=\int^t_{t_0}\mu dt'+L$ with $\mu$
the velocity of the solitons, and the parameters $t_0$ and
$\varphi_0$ are respectively the time and phase at position $z=L$.
The dimensionless parameter $\eta$ characterizes the "grayness"
with $\eta=1$ corresponding to a "dark-soliton" with a $100\%$
density depletion. With $\eta\neq 1$, the solitons travel at a
sound velocity $\mu=c_s(1-\eta^2)^{1/2}+\hbar k/m$ where
$c_s=\sqrt{4\pi|\Phi_2|^2\hbar^2a_{22}/m^2}$ is the maximum values
of the speed in rest frame of background pulse.

The influence of frequency chirp effect on the gray-soliton
dynamics can be studied within the framework of perturbation
theory \cite{12,13}. Denoting by $\alpha=\cos^{-1}\eta$ the
soliton phase angle, and introducing the new variables $d{\cal
T}\approx U_{22}|\Phi_2|^2dt$ and
$d\xi\approx\sqrt{mU_{22}}|\Phi_2|dz$, we obtain the equation of
soliton phase angle as$: \frac{d\alpha}{d{\cal
T}}=\frac{1}{2}\cos^2\alpha\int^{+\infty}_{-\infty}\frac{d{\cal
T}}{\cosh^2{\cal Z}}(\frac{1}{|\Phi_2|}
\frac{\partial|\Phi_2|}{\partial {\cal Z}}) $, where
$\Phi_2=\Phi_2({\cal Z},{\cal T})$ is in the moving frame (with
the solitons), ${\cal Z}=\cos\alpha({\cal T})\bigl[\xi-\xi_0({\cal
T})\bigl]$ with $\xi_0({\cal T})=\sqrt{mU_{22}}|\Phi_2|z_0({\cal
T})$ being the centra position of the soliton. The above formula
clearly shows that the gray-soliton evolution is independent on
the variations of background pulse phase, therefore the frequency
chirp of the background wave have no influence on the solitons.
This indicates that the soliton atom laser can perfectly maintain
its dynamical properties in the propagation, at least for present
configuration.

In addition, for a Gaussian input probe pulse, i.e. the amplitude of
the background wave decreases according to the maximum amplitude of
a dispersively spreading Gaussian pulse shown in Fig.2(a)
(dot-dashed curve), a second-order soliton can split into two
solitons propagating at opposite directions (Considering the
two-soliton solution, which start from the same point and move in
the opposite direction). This phenomena was shown in Fig.3(a,b),
where $\bar z=z-vt$ is the variable in the rest frame of the
background wave. With the decreasing of background amplitude, the
velocity of solitons can be slowed in proportion to the background
intensity (for $c_s\propto|\Phi_2|$), while the spatial width
becomes wider (Fig.3(a)). Due to the phase angle evolution, however,
these effects may be compensated (Fig.3(b)).

It deserves to note that the subtle problem of quantum depletion
is omitted in the above formalism. As is shown in recent
experiments \cite{depletion}, the quantum fluctuation $\hat\psi_2$
grows after the dark solitons are created by the condensate, it
may thus affect the dynamical properties of the created solitons.
Here, to gave an approximate but intuitive description, we denote
the state function of non-condensate atoms as:
$\psi_2(z,t)=\langle\hat\psi\rangle=R(z,t)\Psi_2(z,t)$, where
$R(z,t)$ is the ratio function. The bosonic field $\Psi_2$ would
then be transformed as: $\Psi_2\rightarrow S(z,t)\Psi_2$ with
$|R|^2+|S|^2=1$. Therefore one can steadily obtain:
\begin{equation}
\frac{d\alpha}{d{\cal
T}}=\frac{1}{2}\cos^2\alpha\int^{+\infty}_{-\infty}\frac{d{\cal
T}}{\cosh^2{\cal Z}}\biggr(\frac{1}{|\Phi_2|}
\frac{\partial|\Phi_2|}{\partial {\cal Z}}+\frac{1}{|S|}
\frac{\partial|S|}{\partial {\cal Z}}\biggr)
\end{equation}
which describes the effect of the quantum depletion on the
dynamical properties of the gray solitons. Of course, if one seeks
to quantitatively describe this effect, the general form of the
ratio function $R(z,t)$ should be calculated by developing further
theoretical technique in future works which in fact, as far as we
know, still remains an intriguing and challenging issue in present
literatures \cite{qd}.

Summing up, we have proposed a scheme to obtain a soliton atom
laser with nonclassical atoms based on the quantum state transfer
process from light to a dense atomic beam. In presence of the
intrinsic nonlinear atomic interactions, the quantum state
transfer mechanism is confirmed here from the photons to the
atomic beam. An atomic frequency chirp effect is revealed in the
transfer process, and it is shown to have no influence on the
dynamics of created atomic gray solitons. Finally, some
interesting dynamical properties of gray solitons may happen such
as the splitting of solitons for a Gaussian input probe field, and
the possible effect of quantum depletion is also briefly
discussed. As far as we know, this scheme firstly provides the
possibility to design and probe a soliton atom laser with
nonclassical atoms and its intriguing properties in next
generation of EIT-based experiments. In addition, the nonclassical
properties of the solitons deserves further study, since quantum
states of these solitons can be manipulated with present
technique. Also, our method can be readily extended to analyze the
interested phenomena of other physical systems like an atomic beam
with double-$Lambda$ 4-level configuration \cite{7,14} or even a
fermionic atom laser beam \cite{15}, in which some interesting new
effects should be expected. While much works are needed to clarify
the effects of practical experimental circumstances like the
non-condensed atomic noise, the optimized formation conditions for
the atomic solitons and its dynamical stability problems, our
scheme here should
readily lend itself to such studies. \\

We thank Xin Liu for helpful discussion. This work is supported by
NUS academic research Grant No. WBS: R-144-000-071-305, and by NSF
of China under grants No.10275036 and No.10304020.

%\section{Results}
%\section{Conclusions}
%\indent
%ÕýÎÄ

%%%%%%%%%%%%%%%%%%%%%%%%%%%%%%%%%%%%%%%%%%%%%

%%%%%%%%%%%%%%%%%%%%%%%%%%%%%%%%%%%%%%%%%%%%%
\noindent\\  \\

\begin{figure}[ht]
\includegraphics[width=0.6\columnwidth]{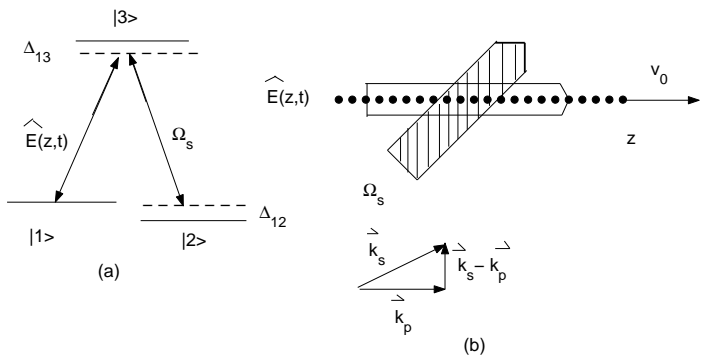}
\caption{(a)Beam of double $\Lambda$ type atoms coupled to a
control field and a quantized probe field. (b)To minimize effect
of Doppler-broadening, geometry is chosen such that
$(\vec{k}_{p}-\vec{k}_{s})\cdot\vec{e}_z\approx0$.} \label{}
\end{figure}

\begin{figure}[ht]
\includegraphics[width=0.8\columnwidth]{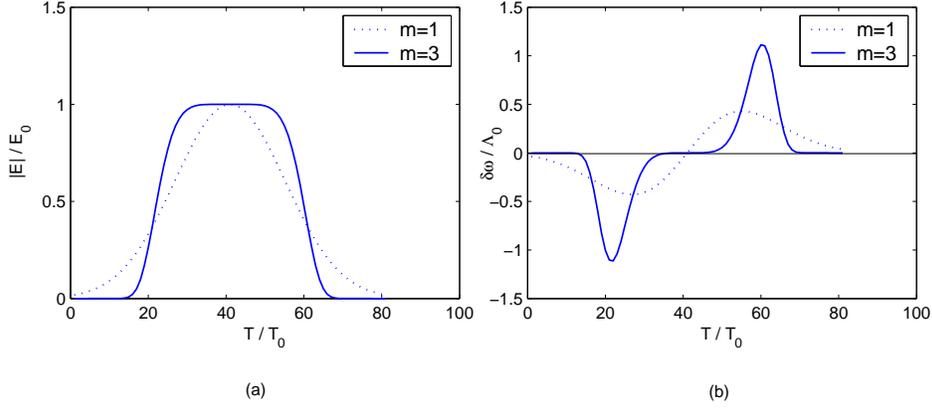}
\caption{(color online) (a). Envelop of the input probe light,
where $E={\cal E}(T)$ and $E_0=|{\cal E}(0)|$ is the maximum of
the amplitude. The Gaussian probe field ($m=1$, dot-dashed curve),
and super Gaussian probe field ($m=3$, solid curve). (b).
Frequency chirp of the output atom laser under the case of
Gaussian input probe field (for $m=1$) and super Gaussian field
(for $m=3$).}\label{}
\end{figure}

\begin{figure}[ht]
\includegraphics[width=0.8\columnwidth]{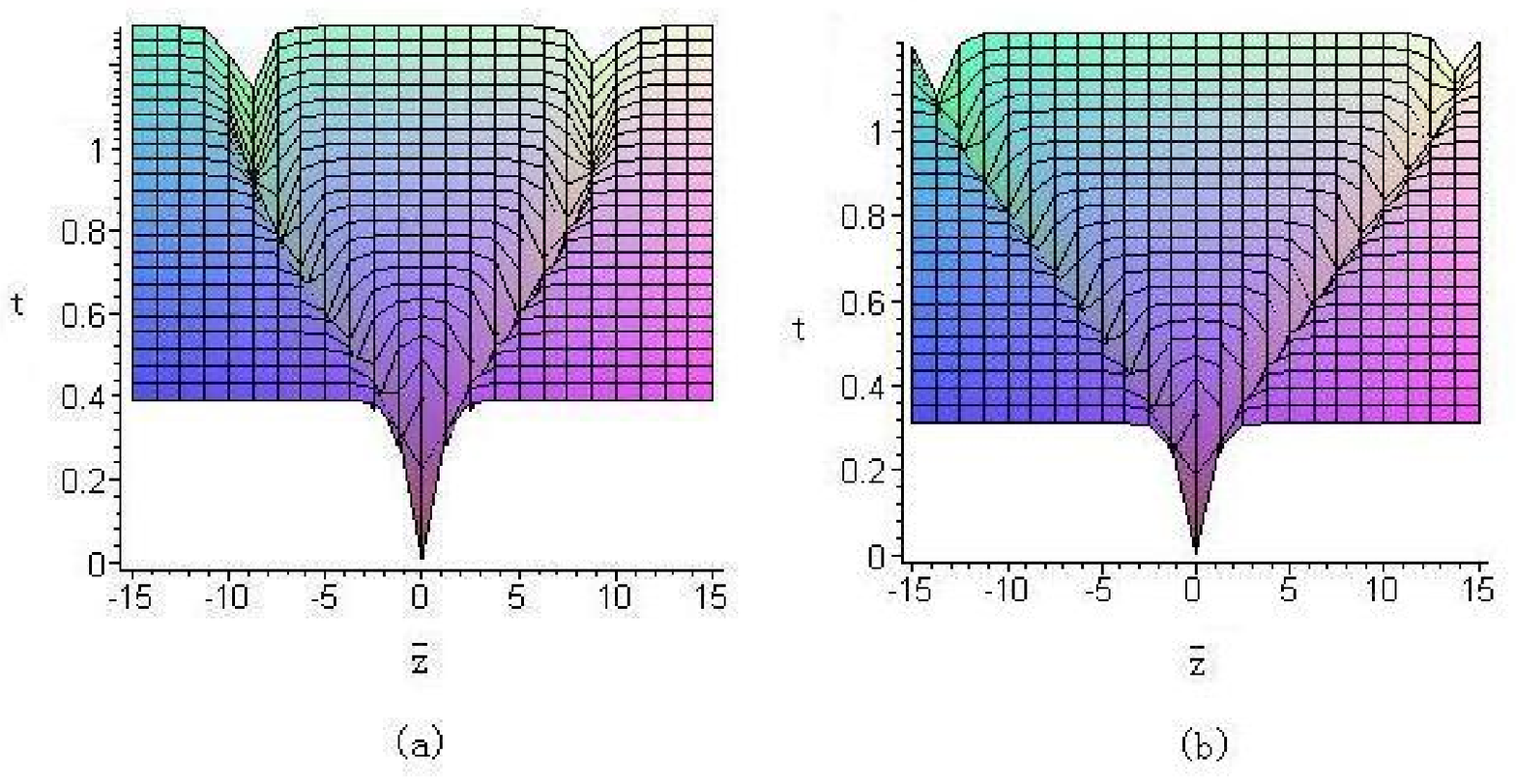}
\caption{(color online) Second-order soliton splits into two
solitons under the influence of background amplitude decreasing
according to the maximum amplitude of a dispersively spreading
Gaussian pulse shown in fig.2(a) (dot-dashed curve). (a) Under a
constant parameter $\eta=0.5$; (b) The evolution of soliton phase
angle is considered.}\label{}
\end{figure}


\begin{thebibliography}{99}


\bibitem{atom laser} M.-O. Mewes et al., Phys.Rev.Lett., 78,582(1997); M. R. Andrews et al., Science 275,637(1997).

\bibitem{interferometer} J. Baudon et al., J. Phys. B: At. Mol. Opt. Phys. 32, R173 (1999);
           A. Peters et al., Metrologia 38, 25 (2001); Y. Shin et al., Phys. Rev. Lett. 92,050405(2004).

\bibitem{noise} C. Santarelli et al. Phys. Rev. Lett., 82,4619(1999); S. F. Huelga et al., Phys. Rev. Lett., 79,3865(1997);
                A. Kuzmich et al., Phys. Rev. Lett., 85,1594(2000).

\bibitem{Drummond} P. D. Drummond et al., Phys. Rev. A 63,053602(2001).

\bibitem{Zhang} W. P. Zhang et al., Phys. Rev. Lett. 72, 60 (1994).

\bibitem{Leboeuf} P. Leboeuf et al, Phys. Rev. A 68, 063608 (2003).

\bibitem{Solitons} K. Porsezian et al, Eds. Optical Solitons (Springer-Verlag, Berlin Heidelberg 2003).

\bibitem{Opinterferometer} M. J. Werner, Phys. Rev. Lett. 81, 4132(1998);
                           M. Fiorentino et al., Phys. Rev. A 64, R031801 (2001).

\bibitem{Feshbach} M. W. Zwierlein, et al., Phys. Rev. Lett. 92, 120403
(2004); and references therein. Note that the Feshbach resonance
point was remarkably proved to be most stable in this new
experimental work. For previous different works, see e.g., E. A.
Donley et al., Nature 412, 295 (2001).
\bibitem{1} L.V.Hau et al., Nature (London) 397,594(1999);
            C. Liu et al., Nature (London) 409,490(2001);
            D. F. Phillips et al., Phys. Rev. Lett. 86,783(2001);
            M. Bajcsy et al., Nature (London) 426,6967(2003).
\bibitem{2} M. D. Lukin et al., Phys. Rev. Lett. 84, 1419(2000);
            M. D. Lukin et al., Phys. Rev. Lett. 84, 4232(2000).
\bibitem{3} Y. Wu et al., Phys. Rev. A 67, 013811 (2003);
            Y. Wu et al. Deng, Opt. Lett. 29, 2064 (2004);
            X. J. Liu et al, Phys. Rev. A, 70, 055802 (2004).
\bibitem{4} M. Fleischhauer and M.D.Lukin, Phys. Rev. Lett. 84, 5094
            (2000); M.Fleischhauer and M.D.Lukin, Phys. Rev. A 65,022314 (2002);
            C. P. Sun et al, Phys. Rev. Lett. 91,147903 (2003).
\bibitem{5} S. E. Harris et al., Phys. Rev. A 46, R29 (1992);
            M. O. Scully et al, Quantum Optics (Cambridge University Press, Cambridge 1999).

\bibitem{6} M. Fleischhauer and S. Q. Gong, Phys. Rev. Lett. 88, 070404 (2002)

\bibitem{7} X. J. Liu et al, Phys. Rev. A, 70,015603(2004).

\bibitem{9} S. A. Morgan et al, Phys. Rev. A 55,4338(1997);
             W. P. Reinhardt et al, J. Phys. B 30,L785(1997).
\bibitem{10} T. Busch et al, Phys. Rev. Lett., 84,2298(2000).
\bibitem{11} J. Denschlag et al., Science, 287,97(2000).

\bibitem{chirp} E. N. Tsoy et al., Phys. Rev. E 62,2882(2000); M. Desaix, Phys. Rev. E 65,056602(2002).
                S. Zamith et al., Phys. Rev. Lett., 87,033001(2001); G. P. Djotyan et al., Phys. Rev. A 68,053409(2003).
\bibitem{eq} M. Ma\v{s}alas and M. Fleischhauer, Phys. Rev. A, 69(RP), 061801(2004).
\bibitem{potential} G. Juzeli\={u}nas and P. \"{O}hberg, cond-mat/0402317(2004).
\bibitem{GPG} F. Dalfovo et al, Rev. Mod. Phys. 71,463(1999).
\bibitem{12} Y. S. Kivshar et al., Rev. Mod. Phys., 61,763(1989).
\bibitem{13} Y. S. Kivshar et al., Phys. Rev. E, 49,1657(1994).
\bibitem{depletion} S.Burger et al., Phys. Rev. Lett. 83,5198(1999).
\bibitem{qd} J. Dziarmaga et al., Phys. Rev. A 66,043615(2002);
             C. K. Law, Phys. Rev. A 68,015602(2003); B. Damski, Phys. Rev. A
             69,043610(2004).
\bibitem{14} X. J. Liu et al, arXiv:quant-ph/0403171(2004).
\bibitem{15} C. Regal et al, Phys. Rev. Lett., 92,040403(2004).
\end{thebibliography}
\end{document}